\documentclass[reprint, aps, prl, superscriptaddress, twocolumn]{revtex4-2}

\usepackage[utf8]{inputenc}
\usepackage{amsmath}
\usepackage{amssymb}
\usepackage{graphicx}
\usepackage{comment}
\usepackage{soul}
\usepackage[allcolors=blue]{hyperref}
\usepackage[dvipsnames]{xcolor}

\begin{document}

\title{Genuine Tripartite Strong Coupling\\ in a Superconducting-Spin Hybrid Quantum System}

\author{Yingqiu Mao}
\email{myingqiu@ustc.edu.cn}
\affiliation{Hefei National Research Center for Physical Sciences at the Microscale and School of Physical Sciences, University of Science and Technology of China, Hefei 230026, China}
\affiliation{Shanghai Research Center for Quantum Science and CAS Center for Excellence in Quantum Information and Quantum Physics, University of Science and Technology of China, Shanghai 201315, China}

\author{Han-Yu Ren}
\affiliation{Hefei National Research Center for Physical Sciences at the Microscale and School of Physical Sciences, University of Science and Technology of China, Hefei 230026, China}
\affiliation{Shanghai Research Center for Quantum Science and CAS Center for Excellence in Quantum Information and Quantum Physics, University of Science and Technology of China, Shanghai 201315, China}

\author{Zi-Yi Liu}
\affiliation{Hefei National Research Center for Physical Sciences at the Microscale and School of Physical Sciences, University of Science and Technology of China, Hefei 230026, China}
\affiliation{Shanghai Research Center for Quantum Science and CAS Center for Excellence in Quantum Information and Quantum Physics, University of Science and Technology of China, Shanghai 201315, China}

\author{Yi-Zheng Zhen}
\affiliation{Hefei National Research Center for Physical Sciences at the Microscale and School of Physical Sciences, University of Science and Technology of China, Hefei 230026, China}
\affiliation{Shanghai Research Center for Quantum Science and CAS Center for Excellence in Quantum Information and Quantum Physics, University of Science and Technology of China, Shanghai 201315, China}

\author{Tao Rong}
\affiliation{Hefei National Research Center for Physical Sciences at the Microscale and School of Physical Sciences, University of Science and Technology of China, Hefei 230026, China}
\affiliation{Shanghai Research Center for Quantum Science and CAS Center for Excellence in Quantum Information and Quantum Physics, University of Science and Technology of China, Shanghai 201315, China}

\author{Tao Jiang}
\affiliation{Hefei National Research Center for Physical Sciences at the Microscale and School of Physical Sciences, University of Science and Technology of China, Hefei 230026, China}
\affiliation{Shanghai Research Center for Quantum Science and CAS Center for Excellence in Quantum Information and Quantum Physics, University of Science and Technology of China, Shanghai 201315, China}

\author{Zhuo Chen}
\affiliation{Hefei National Research Center for Physical Sciences at the Microscale and School of Physical Sciences, University of Science and Technology of China, Hefei 230026, China}
\affiliation{Shanghai Research Center for Quantum Science and CAS Center for Excellence in Quantum Information and Quantum Physics, University of Science and Technology of China, Shanghai 201315, China}

\author{Zhe-Heng Yuan}
\affiliation{Hefei National Research Center for Physical Sciences at the Microscale and School of Physical Sciences, University of Science and Technology of China, Hefei 230026, China}
\affiliation{Shanghai Research Center for Quantum Science and CAS Center for Excellence in Quantum Information and Quantum Physics, University of Science and Technology of China, Shanghai 201315, China}

\author{Wen-Hua Qin}
\affiliation{Hefei National Research Center for Physical Sciences at the Microscale and School of Physical Sciences, University of Science and Technology of China, Hefei 230026, China}
\affiliation{Shanghai Research Center for Quantum Science and CAS Center for Excellence in Quantum Information and Quantum Physics, University of Science and Technology of China, Shanghai 201315, China}

\author{Xiaoran Zhang}
\affiliation{Key Laboratory of Quantum Materials under Extreme Conditions in Shandong Province, School of Physics and Physical Engineering, Qufu Normal University, Qufu 273165, China}
\affiliation{Laboratory of High Pressure Physics and Material Science, Advanced Research Institute of Multidisciplinary Sciences, Qufu Normal University, Qufu 273165, China}

\author{Xiaobing Liu}
\affiliation{Key Laboratory of Quantum Materials under Extreme Conditions in Shandong Province, School of Physics and Physical Engineering, Qufu Normal University, Qufu 273165, China}
\affiliation{Laboratory of High Pressure Physics and Material Science, Advanced Research Institute of Multidisciplinary Sciences, Qufu Normal University, Qufu 273165, China}

\author{Ming Gong}
\affiliation{Hefei National Research Center for Physical Sciences at the Microscale and School of Physical Sciences, University of Science and Technology of China, Hefei 230026, China}
\affiliation{Shanghai Research Center for Quantum Science and CAS Center for Excellence in Quantum Information and Quantum Physics, University of Science and Technology of China, Shanghai 201315, China}

\author{Kae Nemoto}
\affiliation{Okinawa Institute of Science and Technology Graduate University, Onna-son, Okinawa 904-0495, Japan}

\author{William J. Munro}
\affiliation{Okinawa Institute of Science and Technology Graduate University, Onna-son, Okinawa 904-0495, Japan}

\author{Johannes Majer}
\email{johannes@majer.ch}
\affiliation{Hefei National Research Center for Physical Sciences at the Microscale and School of Physical Sciences, University of Science and Technology of China, Hefei 230026, China}
\affiliation{Shanghai Research Center for Quantum Science and CAS Center for Excellence in Quantum Information and Quantum Physics, University of Science and Technology of China, Shanghai 201315, China}

\date{\today}

\begin{abstract}
We demonstrate genuine tripartite strong coupling in a solid-state hybrid quantum system comprising a superconducting transmon qubit, a fixed-frequency coplanar-waveguide resonator, and an ensemble of NV$^-$ centers in diamond. Frequency-domain spectroscopy reveals a characteristic three-mode avoided crossing, indicating that single excitations are coherently shared across all three subsystems. At higher probe powers, we observe nonlinear features including multiphoton transitions and signatures of transmon-${}^{14}\mathrm{N}$ nuclear-spin interactions, highlighting the accessibility of higher-excitation manifolds in this architecture. These results establish a new regime of hybrid cavity QED that integrates superconducting and spin degrees of freedom, providing a platform for exploring complex multicomponent dynamics and developing hybrid quantum interfaces.

\end{abstract}
\date{\today}
\maketitle

Hybrid quantum systems combine complementary physical platforms to realize functionality unattainable in any single constituent~\cite{you_2011_atomic,kurizki_2015_quantum,clerk_2020_hybrid}. 
Among these platforms, superconducting circuits equipped with long-lived spin ensembles unite distinct and complementary strengths~\cite{rabl_2006_hybrid,tordrup_2008_holographic,imamolu_2009_cavity}: 
superconducting qubits offer strong nonlinearities and fast control~\cite{blais_2021_circuit}, while solid-state spin ensembles provide long coherence times and large collective coupling strengths~\cite{dutt_2007_quantum,bienfait_2016_controlling}. Integrating these elements in a single device enables coherent conversion, storage, and manipulation of quantum information across disparate degrees of freedom and forms a promising route toward scalable hybrid quantum networks~\cite{kurizki_2015_quantum,xiang_2013_hybrid,julsgaard_2013_quantum,zhu_2014_observation,kersten_2023_triggered}.

A key step toward such architectures is the realization of coherent interactions involving multiple quantum components. 
Experimental progress has been established in 
strong coupling between microwave resonators and spin ensembles, such as color centers~\cite{schuster_2010_highcooperativity,kubo_2010_strong,amsuss_2011_cavity,ranjan_2013_probing,grezes_2014_multimode,angerer_2018_superradiant}, ferromagnets~\cite{zhang_2014_strongly,goryachev_2014_high,tabuchi_2014_hybridizing}, and doped ions~\cite{schuster_2010_highcooperativity,probst_2013_anisotropic},
and between spin ensembles and superconducting qubits~\cite{zhu_2011_coherent,kubo_2011_hybrid,saito_2013_towards,tabuchi_2015_coherent}, based on which the feasibility of quantum state transfer, storage, and retrieval is tested.
These efforts have largely relied on sequential two-body interactions~\cite{zhu_2011_coherent,kubo_2011_hybrid,saito_2013_towards} or mediator-based coupling protocols~\cite{kubo_2011_hybrid}. 
To go beyond these paradigms and overcome scalability limitations inherent in pairwise coupling, the realization of triple resonance among resonator, superconducting qubits, and spin ensemble is of critical importance.
In particular, tripartite coupling supports hybridized polaritons that simultaneously involve photonic, nonlinear, and collective-spin character, and provides a powerful testbed for studying multicomponent cavity-QED phenomena.

In this work, we realize genuine tripartite strong coupling in a system consisting of a fixed-frequency bus resonator, a superconducting transmon qubit, and an ensemble of negatively charged nitrogen-vacancy (NV$^-$) centers in diamond.
By tuning the transmon into resonance with both the resonator and the spin ensemble, we access a regime in which all coupling rates exceed relevant decoherence rates. Frequency-domain spectroscopy reveals a three-mode avoided crossing with well-resolved polaritonic branches, confirming that excitations are coherently shared across all subsystems. At higher probe powers, we observe nonlinear responses, including multiphoton transitions and hyperfine-assisted features, which demonstrate the ability of this architecture to access excitation manifolds beyond the simple one-excitation picture. Together, these results establish a unified solid-state platform that integrates superconducting, photonic, and spin degrees of freedom within a single strongly coupled device.

\begin{figure*}[!thb]
\includegraphics[width=\textwidth]{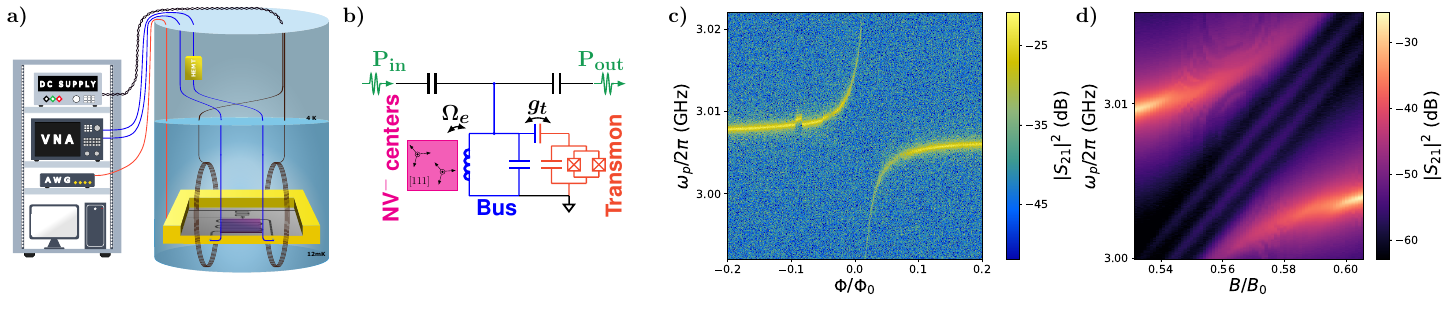}
\caption{\label{fig:setup} 
(a) Schematic illustration of the setup, including control and measurement systems, and the hybrid quantum system. The chip is mounted in the mixing chamber of a dilution refrigerator operating at about $12$ mK. Input lines are equipped with cryogenic attenuators and filters to suppress thermal noise, while the output signal is routed through a circulator to cryogenic and then room temperature amplifier before detection by a vector network analyzer.
A magnetic field is applied using a three-dimensional Helmholtz coil, and the transmon frequency is tuned by a DC bias line.
(b) Circuit diagram of the hybrid quantum system.
(c) Bus-transmon transmission spectra as DC bias is changed in the diamond-loaded case. 
(d) Bus-ensemble transmission spectra as the NV$^{-}$ ensemble is tuned into resonance with the bus mode via adjusting magnetic field tuning, when the transmon is decoupled.
Additional weak peaks between the polaritons arise from the hyperfine structure of NV$^-$ centers associated with the intrinsic $^{14}\mathrm{N}$ nuclear spins.
}
\end{figure*}

Our hybrid system, illustrated in Figs.~\ref{fig:setup}(a-b), comprises a superconducting transmon qubit, a fixed-frequency coplanar-waveguide (CPW) bus resonator, and an ensemble of NV$^{-}$ centers. The transmon is capacitively coupled to the bus resonator, with its transition frequency tunable via a DC flux bias. The diamond containing the NV$^-$ ensemble is positioned at the antinode of the magnetic field of the resonator to maximize coupling. By applying an external magnetic field, the NV$^-$ transition frequencies can be tuned through the Zeeman effect, allowing controlled resonance with the bus resonator.
The system's spectral response is probed using frequency-domain microwave transmission measurements under weak excitation conditions.

The individual couplings between the bus resonator and each subsystem are shown in Figs.~\ref{fig:setup}(c-d), with detailed characterizations given in Sec.~I of Supplemental Material (SM)~\footnote{See Supplemental Material for additional data and analysis.}. With the diamond loaded, the bus frequency is $3.007$ GHz and has linewidth (FWHM) $\kappa/2\pi = 0.171$ MHz. The bus-transmon spectroscopy is measured in the absence of an external magnetic field, which ensures the frequency of the ensemble is far detuned from the bus and effectively decoupled. A clear avoided crossing is observed in Fig.~\ref{fig:setup}(c), corresponding to a bus-transmon coupling strength of $g_{\mathrm{t}}/2\pi = 17.490\pm 0.208$ MHz.
The probe power is then increased $55$ dB to measure the bus-ensemble spectroscopy, so that the bus-transmon coupling is suppressed while the bus-ensemble system is still in the linear-response regime. By optimizing the angle and amplitude of the external magnetic field, the NV$^-$ ensemble is tuned to resonance with the bus mode.  From the resonance spectra (Fig.~\ref{fig:setup}d), the normal-mode splitting suggests a collective coupling of the bus-ensemble of $\Omega_{e}/2\pi = 6.597\pm0.092$ MHz.

\begin{figure}[b]
    \includegraphics[width=1\columnwidth]{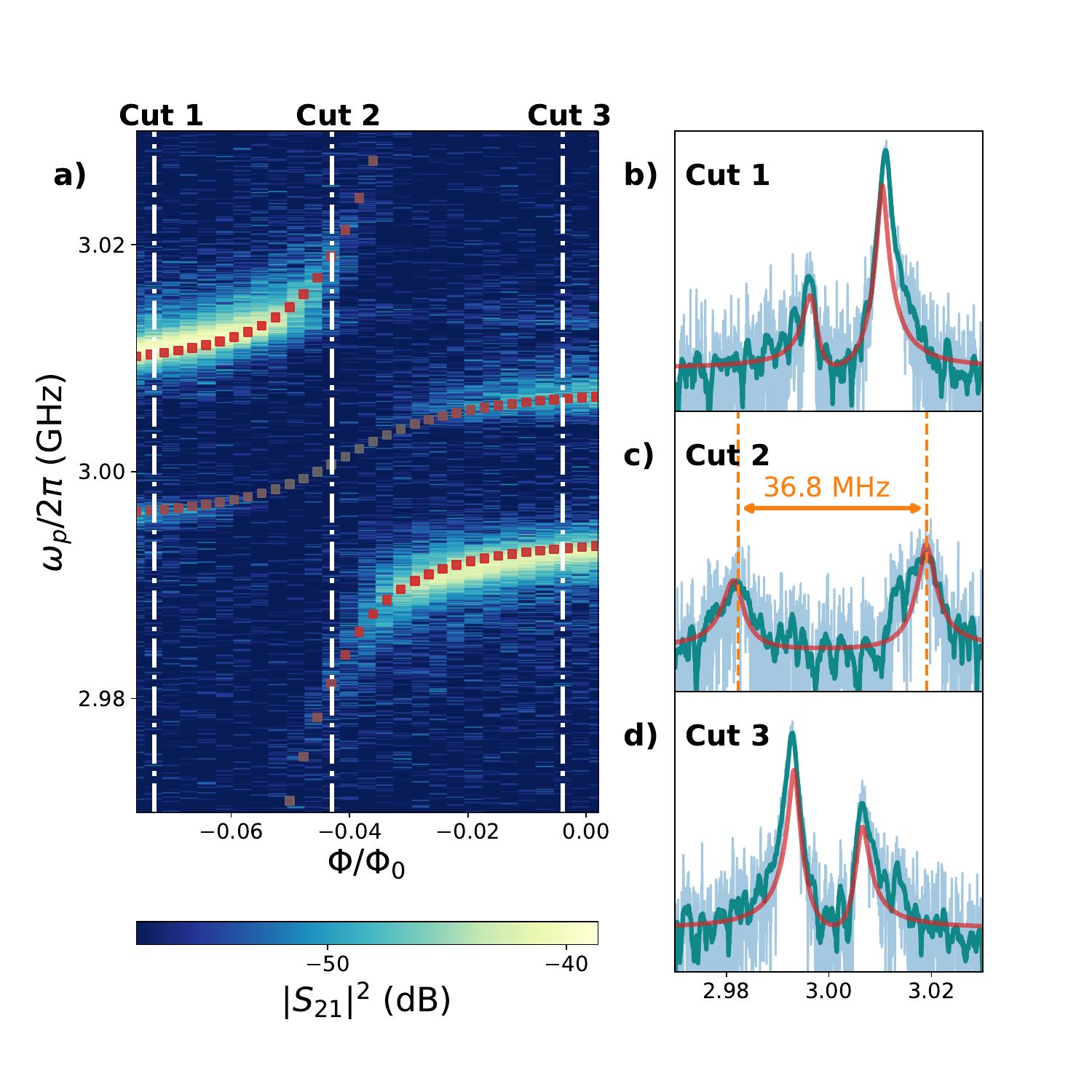}
    \caption{\label{fig:hybrid-spectr} 
    (a) Transmission spectroscopy of the hybrid system at a probe power of $-70$ dBm as the transmon frequency is swept across resonance with the bus and NV$^-$ ensemble.
    Red dotted lines: transition frequencies of the hybrid system under one-excitation approximation, with color shade indicating the proportion of photons in the resonator.
    (b-d) Transmission coefficients of three cuts in (a), highlighting the hybridized polaritonic modes in three resonance cases.
    Pale-blue lines: measured data. Green lines: processed data for guiding eyes. Red lines: theoretical simulations.
    }
\end{figure}

Reducing the probe power to single-excitation level places the system in the regime where genuine tripartite coupling can be observed. The transmon frequency is swept and the transmitted signal is recorded, yielding the spectrum shown in Fig.~\ref{fig:hybrid-spectr}. It reveals the system smoothly transforming from nearly a two-mode bus-ensemble hybridization (Cut 1), to a genuine three-mode hybrid at triple resonance (Cut 2), and back again (Cut 3) as the transmon detunes. Specifically, when the transmon frequency approaches the resonant frequency of the bus and NV$^-$ ensemble, the bus-ensemble polaritons begin to dress the transmon states. As the hybridized energy levels reorganize, one bright polariton shifts outward with its frequency moving away, the other polariton loses resonator character and becomes dark in transmission, and simultaneously a new bright polariton emerges. Due to the increased effective coupling between the bus and collective transmon-spin subsystem, on resonance~\footnote{The resonant frequency is slightly shifted to $3.0012$ GHz due to multiple adjustments of magnetic field and fridge temperature throughout the experiment.}, the two bright-mode polaritons exhibit avoided crossing with a minimal separation approximately $36.8$ MHz, exceeding the splitting when bus couples to transmon alone (approximately $34.2$ MHz) or to NV$^-$ ensemble alone (approximately $13.2$ MHz).
These pronounced spectral features confirm coherent coupling among all three subsystems, and the measured mean bright-polariton linewidth on resonance, i.e., $\Gamma_h$, is about $6.0$ MHz (see details in Sec.~III of SM), which further confirms the hybrid system is in the strong-coupling regime.

To elucidate the experimental observations further, a simplified Hamiltonian capturing the essential three-mode interactions is considered. The resonator field is described as a single harmonic mode, while the transmon is approximated as a qubit~\cite{blais_2021_circuit}. The spin ensemble is strongly coupled to the bus resonator, and in the low-excitation limit, it can be approximated as a harmonic oscillator with the collective mode~\cite{kurucz_2011_spectroscopic,diniz_2011_strongly}. The total Hamiltonian can be written as
\begin{align}
H &= \hbar\omega_ra^\dagger a+\hbar \omega_t\frac{\sigma_t^z}{2} + \hbar \omega_s S^\dagger S \nonumber \\
& + \hbar g_t (a^\dagger \sigma_t^- +a \sigma_t^+) + \hbar\Omega_e (a^\dagger S +a S^\dagger),\label{eq:Hamiltonian_simple}
\end{align}
where $a,S$ ($a^\dagger, S^\dagger$) are the annihilation (creation) operators for the bus mode and the collective ensemble mode, respectively, $\sigma_t^{z,\pm}$ are Pauli operators for the transmon qubit, $\omega_r,\omega_t,\omega_s$ are the transition frequencies of the resonator, transmon, and spin ensemble respectively. 

Diagonalizing Eq.~\eqref{eq:Hamiltonian_simple} in the bright-dark basis yields three hybridized eigenmodes, with details provided in Sec. II of SM~\cite{Note1}. Using the measured parameters of the individual systems, theoretical simulations of transition frequencies from ground state to the eigenmodes (red dotted lines) agree well with positions of three polaritons measured in experiment, as shown in Fig.~\ref{fig:hybrid-spectr}(a). Specifically, at triple resonance $\omega_r=\omega_t=\omega_s$, the bright and dark eigenmodes are 
\begin{equation}
|1_\pm\rangle=\frac{|1\rangle|g\rangle|G\rangle \pm |0\rangle|B\rangle}{\sqrt{2}},\quad
|1_D\rangle=|0\rangle|D\rangle,
\end{equation}
with eigenenergies $E_\pm = \hbar(\omega_r\pm\Omega_h)$ and $E_D=\hbar\omega_r$, respectively. Here, the symmetric and asymmetric states
\begin{equation}
|B\rangle=\frac{g_t|e\rangle|G\rangle+\Omega_e|g\rangle|E\rangle}{\Omega_h},
|D\rangle=\frac{\Omega_e|e\rangle|G\rangle-g_t|g\rangle|E\rangle}{\Omega_h}.
\end{equation}
where $|0,1\rangle$ represent the zero- and one-photon states in the bus, $|g,e\rangle$ the ground and first levels of transmon, $|G,E\rangle$ the ground and first collective states of NV$^{-}$ ensemble, and $\Omega_h=\sqrt{g_t^2+\Omega_e^2}$ is the hybrid coupling strength between bus and collective transmon-ensemble (transmon qubit and NV$^-$ qubits). 
The experimentally extracted parameters give $\Omega_{h}/2\pi= 18.693\pm0.197$ MHz, which is remarkably close to half of the observed polariton splitting. 

The observation of a dark mode serves as a definitive signature of true tripartite hybridization. If the resonator couples only to the transmon or the spin ensemble (each an effective two-level system exchanging energy with the resonator), no dark mode could form under weak probing. However, as demonstrated above, the transmon and NV$^-$ ensemble constitute an effective three-level quantum system with a common ground state and states $|B\rangle$ and $|D\rangle$. It is the coupling of this three-level system to the bus resonator that enables the dark mode observation. Importantly, the NV$^-$ triplet cannot account for the three-level feature because the applied magnetic field detunes one state far from the resonant levels.

To further characterize the dissipation rate of the system, the transmission coefficient is evaluated using the same model. The transmission $S_{21}$ scales with
\begin{equation}
S_{21}(\omega_p) \propto \frac{1}{\Delta_r + \frac{g_t^2}{\Delta_t} + \Omega_e^2\int d\omega\frac{\varrho(\omega)}{\Delta_\omega}},\label{eq:S21}
\end{equation}
where $\Delta_r = \omega_r - \omega_p - i\kappa/2$, $\Delta_t = \omega_t - \omega_p - i\gamma_t/2$, and $\Delta_\omega = \omega - \omega_p - i\gamma_s/2$ are the complex detunings of the probe frequency $\omega_p$ from the resonator, transmon and spin frequencies, respectively, and $\varrho(\omega)$ is the spectral density profile of the ensemble~\cite{sandner_2012_strong}. 
Using the parameters of the individual subsystem measurements, $|S_{21}|^2$ is simulated with results shown in Figs.~\ref{fig:hybrid-spectr}(b-d), where the line shape agrees with experimental data. 
As the linewidth of the collective transmon-ensemble can be determined through $\Gamma_h=(\kappa+\Gamma_{te})/2$, a cooperativity of $C=4\Omega_h^2/(\kappa\Gamma_{te})\approx 692$ can be estimated, which further supports the strong-coupling feature of the system.

Small discrepancies between the experimental data and theoretical lines remain, which can be attributed to power broadening of the transmon under our spectroscopy conditions and additional dissipation channels introduced by the applied magnetic field (most notably increased loss in the aluminum Josephson junction). Both effects broaden the spectra and degrade the signal-to-noise ratio (SNR) even after extensive averaging, reducing the visibility of fine spectral features. Nonetheless, the pronounced three-mode avoided crossings remain clearly resolved, demonstrating coherent energy exchange and strong tripartite coupling among the bus resonator, transmon, and NV$^{-}$ ensemble.

\begin{figure}
    \includegraphics[width=1\columnwidth,trim=50 130 10 160, clip]{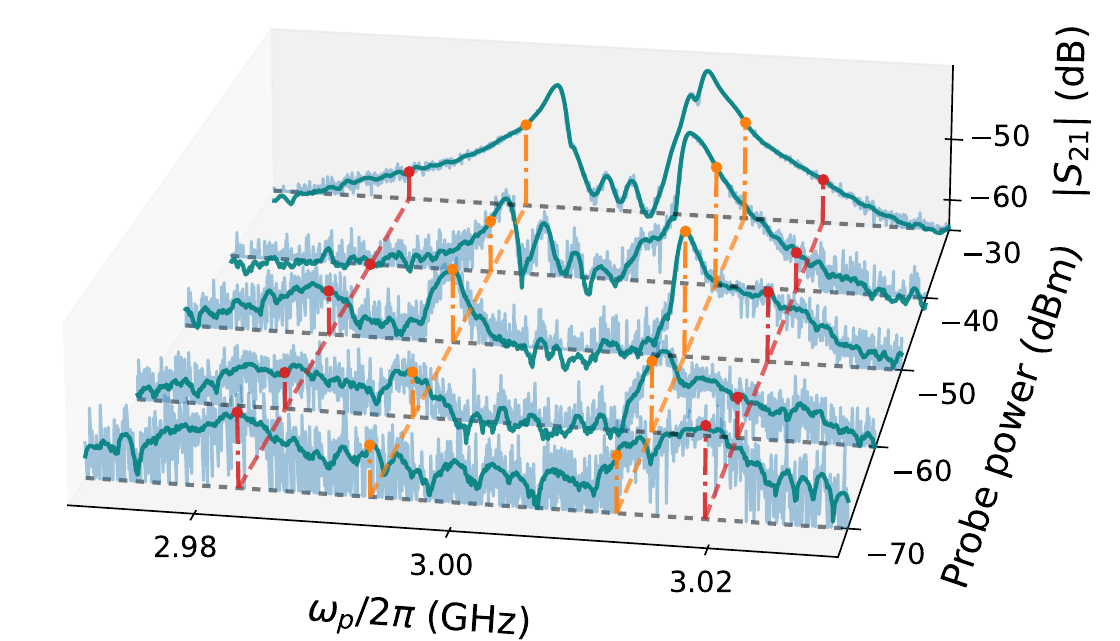}
    \caption{\label{fig:hybrid-nonlinear} 
    Transmission spectra of the hybrid system at various probe powers when the transmon, bus resonator, and NV$^-$ ensemble are tuned into triple resonance. Pale blue lines: raw data. Green lines: processed data for guiding eyes. Red lines: positions of original polariton peaks. Orange lines: positions of additional spectura features.
    }
\end{figure}

Our hybrid quantum system has demonstrated sensitivity to both probe and control fields under low-power conditions, prompting an investigation into the transition from linear to nonlinear behavior with increasing probe power. Consequently, the probe power is varied from $-70$ dBm to $-30$ dBm in $10$ dB increments, with resonance traces compiled in Fig.~\ref{fig:hybrid-nonlinear} and full spectra provided in Sec.~IV of SM~\cite{Note1}. As power increases, the original polariton peaks gradually broaden and submerge, while additional spectral features emerge between them. This is primarily due to AC Stark shift inducing new resonance modes involving the transmon's higher-level transitions and two-photon processes, where the transmon's significant anharmonicity plays a key role~\cite{bishop_2008_nonlinear}. Above $-50$ dBm, the transmon begins to thermalize and gradually decouples as probe power approaches $-30$ dBm, leading to the system's response shifting from three-mode hybridization toward bus-ensemble dominated coupling. This evolution results in a power-dependent response that departs from the linear three-mode model, clearly indicating the excitation of higher manifolds and the breakdown of the low-excitation approximation.

\begin{figure}
    \includegraphics[width=1\columnwidth]{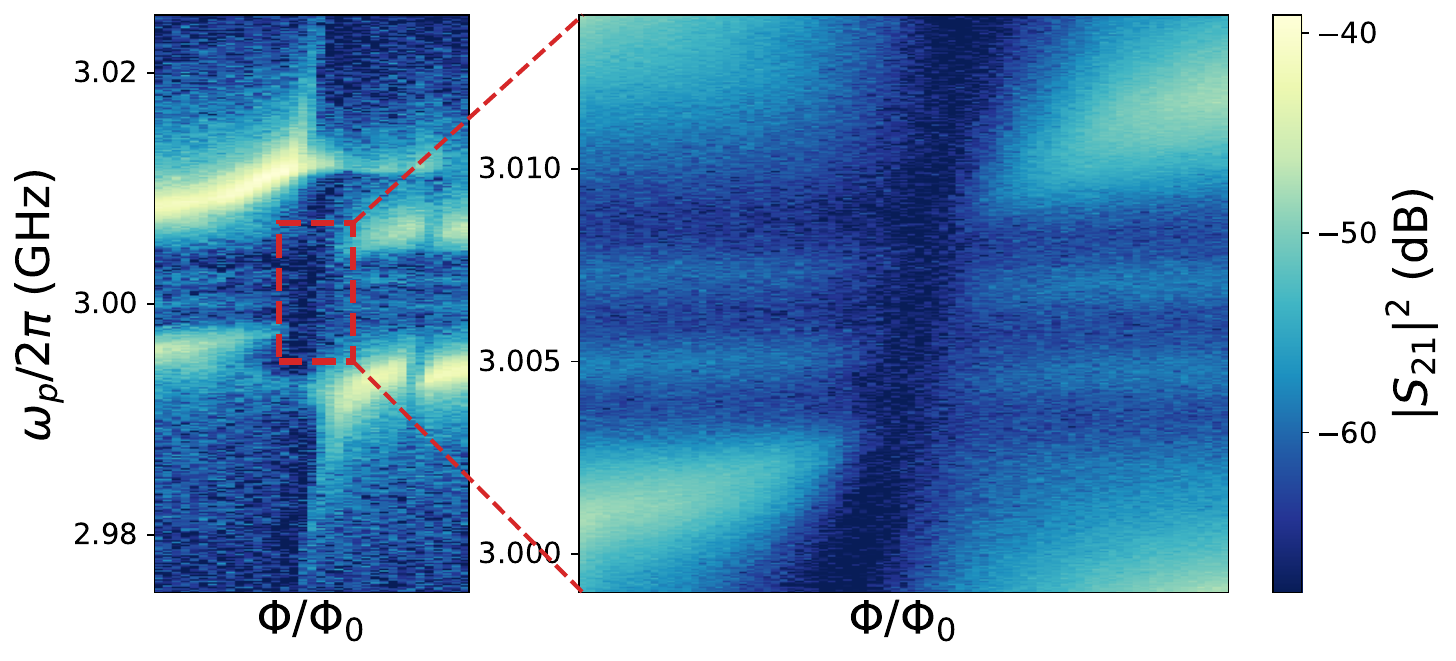}
    \caption{\label{fig:hyper-N14-spectr} 
    Full transmission spectrum of the hybrid system showing fine spectral features arising from the hyperfine structure of the NV$^-$ ensemble. Inset: Magnified view of the highlighted region.
    }
\end{figure}

Furthermore, preliminary spectral features consistent with ${}^{14}\mathrm{N}$ coupling to the transmon are observed at probe power of $-50$ dBm, as shown in the inset of Fig.~\ref{fig:hyper-N14-spectr}.
In addition to power broadening and polariton peak splitting under high probe power, the hyperfine splitting of NV$^-$ centers~\cite{yang_2022_observing} is clearly resolved as two faint horizontal lines near $3$ GHz between the polariton branches.
When the transmon is tuned into resonance, the primary avoided crossing transforms into multiple avoided crossings as fine structures. As analyzed in Sec. V of SM~\cite{Note1}, we confirm that these fine lines are results of coherent coupling among hyperfine-resolved NV$^-$ centers, resonator, and transmon.
Despite limited spectral resolution, such couplings indicate the in-principle feasibility of coherently interfacing the transmon with those nuclear spins for state transfer or storage.

To conclude, our frequency-domain measurements reveal well-resolved polaritonic branches whose structure confirms genuine three-mode hybridization of the transmon, bus resonator, and NV$^{-}$ ensemble.
Under stronger probing, additional resonances corresponding to multiphoton excitations, nonlinear hybrid dynamics, and subtle features consistent with transmon-${}^{14}\mathrm{N}$ nuclear-spin coupling emerge, illustrating the system's accessibility to richer excitation manifolds.
Together, these results provide a direct spectroscopic evidence of coherent three-body coupling in a superconducting-spin cavity-QED device.

These capabilities offer a versatile platform for coherent multicomponent control.
Through the bus resonator, symmetric collective excitations shared by the transmon and NV$^-$ ensemble can be prepared, enabling single-photon-mediated entanglement among physical systems at different scales~\cite{mao_2021_perspective}. 
The transmon tunability also provides controlled access to higher-excitation manifolds and anharmonic spin dynamics, allowing experimental tests of related many-body effects~\cite{lei_2023_manybody}, photon-mediated spin synchronization~\cite{nadolny_2023_macroscopic} and ``quantum-battery'' modes of operation~\cite{kurman_2025_quantum}, establishing a foundation for future investigations of coherence, fidelity, and scaling in multicomponent quantum devices.

\begin{acknowledgments}

We thank valuable discussions with Victor R. Garcia, Qi-Chao Sun, Ya Wang, Yuimaru Kubo, Lev Bishop, and Oscar Dalsten. We also sincerely thank Profs. Jian-Wei Pan, Yu‑Ao Chen, and Xiaobo Zhu for their crucial support to this project.
This work has been supported by the Quantum Science and Technology-National Science and Technology Major Project (No.~2024ZD0301200), National Natural Science Foundation of China (No.~12475028,~12104444), Fundamental Research Funds for the Central Universities (No.~WK9990000159,~WK9990250186), Anhui Provincial Natural Science Foundation (No.~2308085MA26), and China Postdoctoral Science Foundation (No.~2021M693093).
We also acknowledge the Supercomputing Center of USTC for providing computational resources and support.
Y. M. and Y. Z. express sincere gratitude to Profs. Barry C. Sanders and Valerio Scarani for many inspiring discussions and for the constant encouragement to persevere. 

The data that support the findings of this article are not publicly available, and are available from the authors upon reasonable request.

\end{acknowledgments}

\bibliography{SpectroHQS-Exp}

\end{document}


\title{Supplemental Material for\\ Genuine Tripartite Strong Coupling in a Superconducting-Spin\\ Hybrid Quantum System}
\author{Yingqiu Mao}
\email{myingqiu@ustc.edu.cn}
\affiliation{Hefei National Research Center for Physical Sciences at the Microscale and School of Physical Sciences, University of Science and Technology of China, Hefei 230026, China}
\affiliation{Shanghai Research Center for Quantum Science and CAS Center for Excellence in Quantum Information and Quantum Physics, University of Science and Technology of China, Shanghai 201315, China}

\author{Han-Yu Ren}
\affiliation{Hefei National Research Center for Physical Sciences at the Microscale and School of Physical Sciences, University of Science and Technology of China, Hefei 230026, China}
\affiliation{Shanghai Research Center for Quantum Science and CAS Center for Excellence in Quantum Information and Quantum Physics, University of Science and Technology of China, Shanghai 201315, China}

\author{Zi-Yi Liu}
\affiliation{Hefei National Research Center for Physical Sciences at the Microscale and School of Physical Sciences, University of Science and Technology of China, Hefei 230026, China}
\affiliation{Shanghai Research Center for Quantum Science and CAS Center for Excellence in Quantum Information and Quantum Physics, University of Science and Technology of China, Shanghai 201315, China}

\author{Yi-Zheng Zhen}
\affiliation{Hefei National Research Center for Physical Sciences at the Microscale and School of Physical Sciences, University of Science and Technology of China, Hefei 230026, China}
\affiliation{Shanghai Research Center for Quantum Science and CAS Center for Excellence in Quantum Information and Quantum Physics, University of Science and Technology of China, Shanghai 201315, China}

\author{Tao Rong}
\affiliation{Hefei National Research Center for Physical Sciences at the Microscale and School of Physical Sciences, University of Science and Technology of China, Hefei 230026, China}
\affiliation{Shanghai Research Center for Quantum Science and CAS Center for Excellence in Quantum Information and Quantum Physics, University of Science and Technology of China, Shanghai 201315, China}

\author{Tao Jiang}
\affiliation{Hefei National Research Center for Physical Sciences at the Microscale and School of Physical Sciences, University of Science and Technology of China, Hefei 230026, China}
\affiliation{Shanghai Research Center for Quantum Science and CAS Center for Excellence in Quantum Information and Quantum Physics, University of Science and Technology of China, Shanghai 201315, China}

\author{Zhuo Chen}
\affiliation{Hefei National Research Center for Physical Sciences at the Microscale and School of Physical Sciences, University of Science and Technology of China, Hefei 230026, China}
\affiliation{Shanghai Research Center for Quantum Science and CAS Center for Excellence in Quantum Information and Quantum Physics, University of Science and Technology of China, Shanghai 201315, China}

\author{Zhe-Heng Yuan}
\affiliation{Hefei National Research Center for Physical Sciences at the Microscale and School of Physical Sciences, University of Science and Technology of China, Hefei 230026, China}
\affiliation{Shanghai Research Center for Quantum Science and CAS Center for Excellence in Quantum Information and Quantum Physics, University of Science and Technology of China, Shanghai 201315, China}

\author{Wen-Hua Qin}
\affiliation{Hefei National Research Center for Physical Sciences at the Microscale and School of Physical Sciences, University of Science and Technology of China, Hefei 230026, China}
\affiliation{Shanghai Research Center for Quantum Science and CAS Center for Excellence in Quantum Information and Quantum Physics, University of Science and Technology of China, Shanghai 201315, China}

\author{Xiaoran Zhang}
\affiliation{Key Laboratory of Quantum Materials under Extreme Conditions in Shandong Province, School of Physics and Physical Engineering, Qufu Normal University, Qufu 273165, China}
\affiliation{Laboratory of High Pressure Physics and Material Science, Advanced Research Institute of Multidisciplinary Sciences, Qufu Normal University, Qufu 273165, China}

\author{Xiaobing Liu}
\affiliation{Key Laboratory of Quantum Materials under Extreme Conditions in Shandong Province, School of Physics and Physical Engineering, Qufu Normal University, Qufu 273165, China}
\affiliation{Laboratory of High Pressure Physics and Material Science, Advanced Research Institute of Multidisciplinary Sciences, Qufu Normal University, Qufu 273165, China}

\author{Ming Gong}
\affiliation{Hefei National Research Center for Physical Sciences at the Microscale and School of Physical Sciences, University of Science and Technology of China, Hefei 230026, China}
\affiliation{Shanghai Research Center for Quantum Science and CAS Center for Excellence in Quantum Information and Quantum Physics, University of Science and Technology of China, Shanghai 201315, China}

\author{Kae Nemoto}
\affiliation{Okinawa Institute of Science and Technology Graduate University, Onna-son, Okinawa 904-0495, Japan}

\author{William J. Munro}
\affiliation{Okinawa Institute of Science and Technology Graduate University, Onna-son, Okinawa 904-0495, Japan}

\author{Johannes Majer}
\email{johannes@majer.ch}
\affiliation{Hefei National Research Center for Physical Sciences at the Microscale and School of Physical Sciences, University of Science and Technology of China, Hefei 230026, China}
\affiliation{Shanghai Research Center for Quantum Science and CAS Center for Excellence in Quantum Information and Quantum Physics, University of Science and Technology of China, Shanghai 201315, China}

\date{\today}
\maketitle

This Supplemental Material provides the technical background and supporting data for the main text. It is organised in five sections that together give experimental detail, extended characterization, and the theoretical model used to interpret the hybrid tripartite coupling.

\section{Device and Measurement Details}

\begin{figure}[htbp]
\includegraphics[height=8cm]{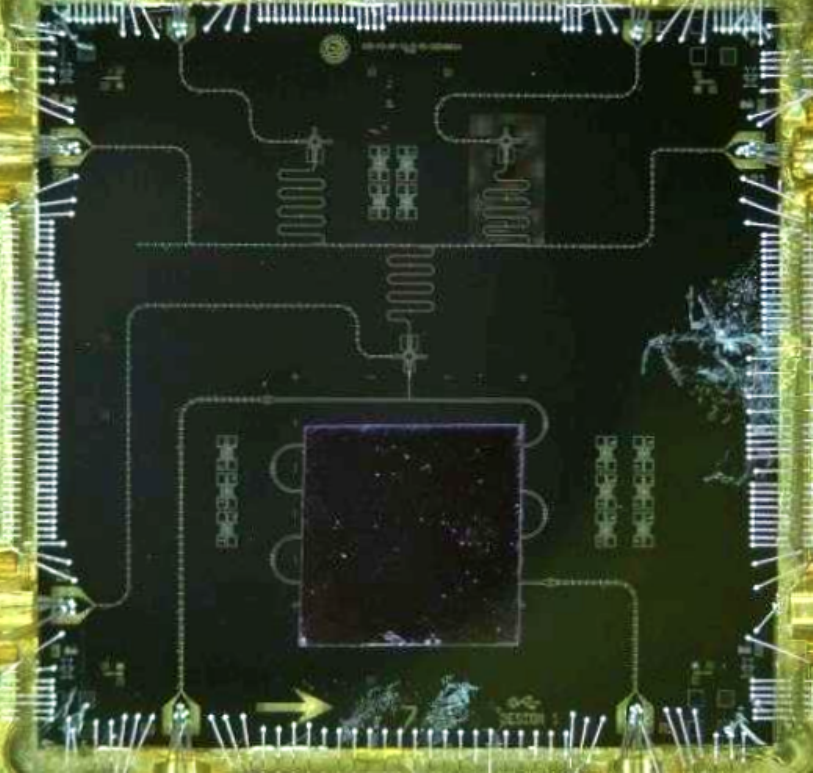}
\caption{\label{fig:chip}
Optical micrograph of the chip consisting of a CPW bus resonator, a transmon, and a diamond placed on it. 
The superconducting circuit is fabricated on a sapphire substrate with a $200$ nm tantalum base layer; the CPW resonator, readout resonator, and control lines are patterned by laser lithography, and the aluminum Josephson junctions are defined by electron-beam lithography.
The diamond crystal, cut along the $(111)$ plane, is the small black square positioned at the center of the CPW resonator.}
\end{figure}

\subsection{Bus resonator}

\begin{figure}[htbp]
    \centering
  \includegraphics[height=5cm]{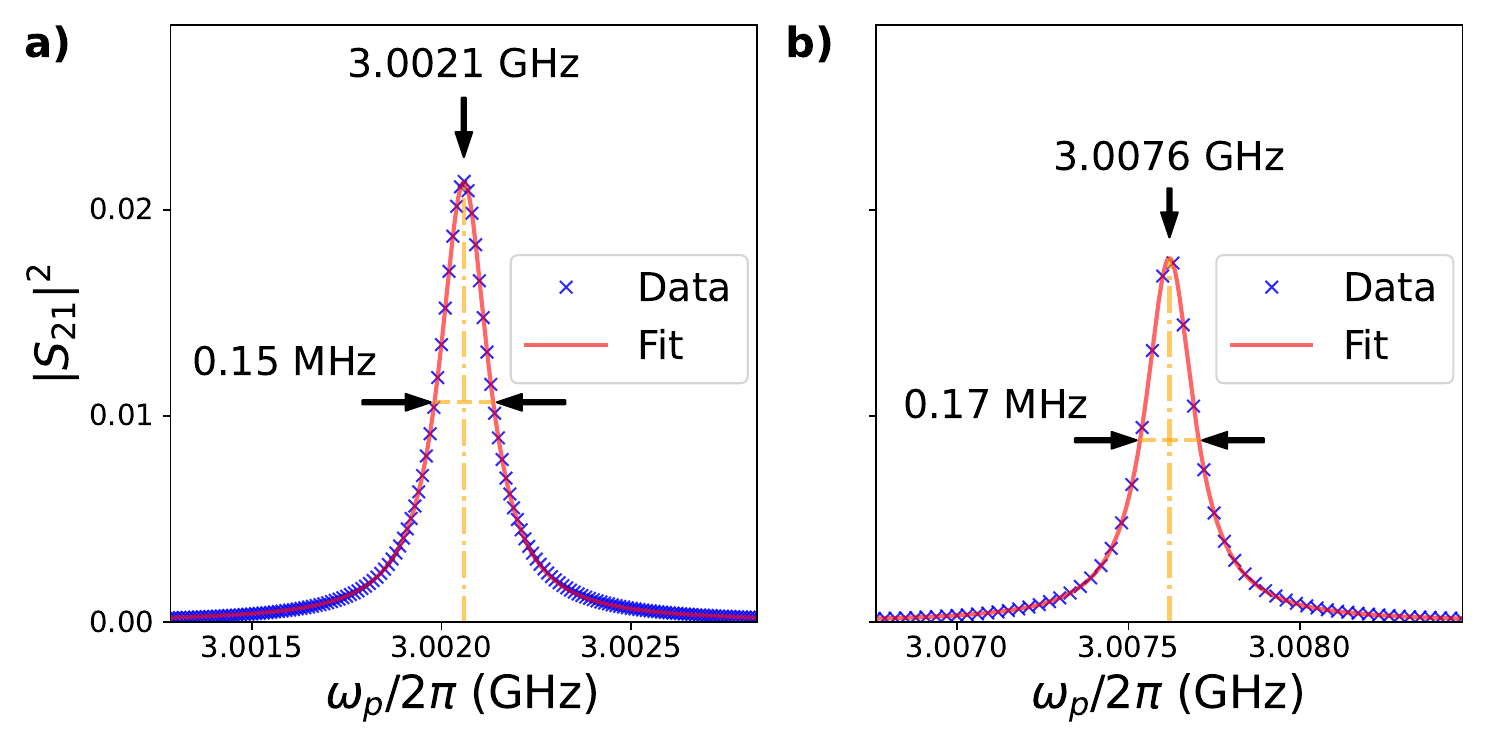}
    \caption{Transmission spectra of the bus resonator with diamond (a) and with diamond plus magnetic field (b). The central frequencies (dash-dot lines) and FWHM values (dashed lines) are indicated for each case.}
    \label{fig:resonator_spectra}
\end{figure}

The bus resonator (see the bottom half of the chip in Fig.~\ref{fig:chip} is characterized through transmission measurements under three distinct conditions: the bare resonator, the resonator with a diamond sample placed on it, and the resonator with the diamond under an external magnetic field applied.
The bare resonator exhibited a resonant frequency of around $3.0961$ GHz, consistent with the designed value. When the diamond is introduced, the frequency shifted to around $3.002058925\pm0.000000045$ GHz , primarily due to the dielectric loading of the diamond sample, and the measured full width at half maximum (FWHM) is around $0.154521\pm0.000116$ MHz (see Fig.~\ref{fig:resonator_spectra}a).
After the superconducting chip has been subjected to external magnetic field, the frequency further adjusted to $3.007619593\pm0.000000029$ GHz, and the FWHM slightly increases to $0.170971\pm0.000086$ MHz (see Fig.~\ref{fig:resonator_spectra}b).
Due to the very small standard deviation, we set $\kappa/2\pi=0.171$ MHz for the later simulation.

\subsection{Transmon}

\begin{figure}[htbp]
    \centering
  \includegraphics[height=5.1cm]{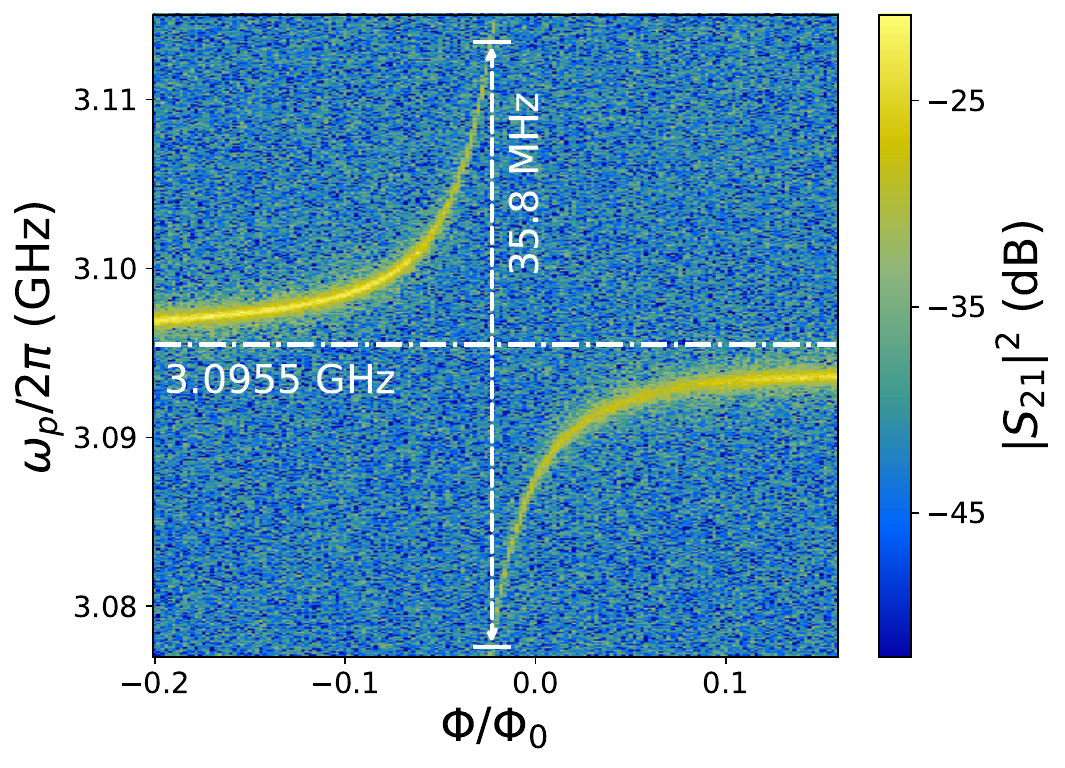}
    \caption{Transmission spectra of the transmon coupled to the bus resonator without the diamond. The resonance frequency (dash-dot line) and coupling strength (dashed line) are indicated.}
    \label{fig:transmon_spectra_wodiamond}
\end{figure}

The transmon qubit (see the center cross-shaped pattern of the chip in Fig.~\ref{fig:chip}) is initially characterized by transmission line measurements, revealing a sweet spot frequency of $3.230$ GHz, and anharmonicity of $203.1$ MHz.
Through time-domain experiments, we determined the coherence times with $T_1\approx 10.3$ $\mathrm{\mu s}$ (energy relaxation time) and $T_2\approx 6.58$ $\mathrm{\mu s}$ (dephasing time).
The coupling between the transmon and the bus resonator is evaluated in the absence of the diamond sample, yielding a coupling strength of around $17.9$ MHz, see Fig.~\ref{fig:transmon_spectra_wodiamond}.
Upon introducing the diamond, the coupling strength remained nearly unchanged, see Fig. 1(c) in the main text.

To quantify the transmon's contribution to the hybrid system, we employ the Jaynes-Cummings (JC) model, which describes the coherent interaction between a two-level system (the transmon) and a quantized cavity mode (the bus resonator).
Under weak probing conditions, the transmission spectrum reveals two distinct polariton states, with their frequencies given by
\begin{equation}
  \omega_{\pm} = \frac{\omega_r+\omega_q}{2} \pm \frac{1}{2}\sqrt{(\omega_r - \omega_q)^2 + 4g_t^2},
\end{equation}
where $\omega_r$ and $\omega_q$ are the resonant frequencies of the bus resonator and transmon, respectively, and $g_t$ is the coupling strength between them.
These spectral peaks exhibit Lorentzian line shapes characterized by center frequencies $\omega_{\pm}$ and FWHM defined by
\begin{align}
  \Gamma_+ &= \kappa \cos^2\theta_1 + \gamma_t \sin^2\theta_1,\\
  \Gamma_- &= \kappa \sin^2\theta_1 + \gamma_t \cos^2\theta_1,
\end{align}	
where $\theta_1$ is the mixing angle satisfying $\tan(2\theta_1) = 2g_t/(\omega_r - \omega_q)$, $\kappa$ is the decay rate of the bus resonator, and $\gamma_t$ is the decay rate of the transmon.

\begin{figure}[bh]
    \centering
    \includegraphics[height=5cm]{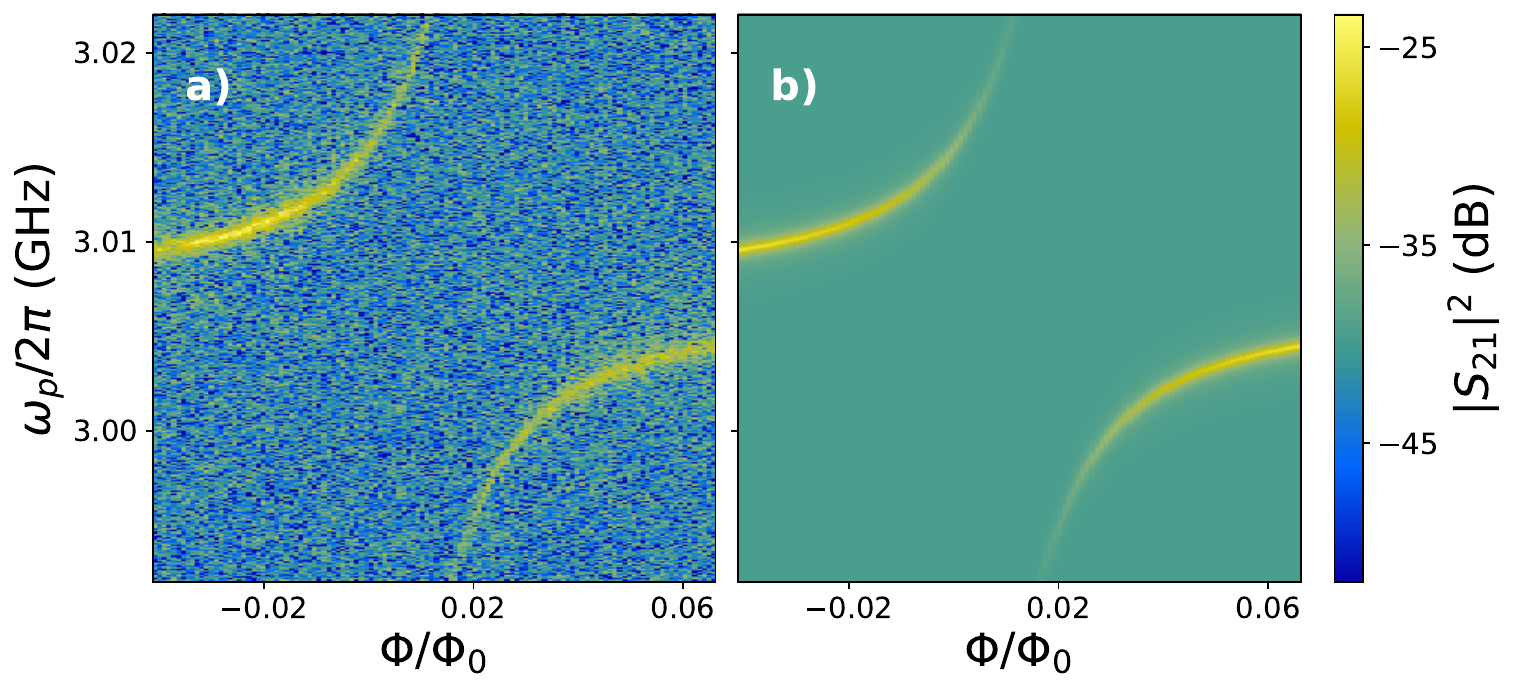}
    \caption{Transmission spectra of the transmon coupled to the bus resonator with the diamond: experimental data (a) and theoretical fit (b).}
    \label{fig:transmon_spectra_fit}
\end{figure}

By fitting the data from Fig.~1(c) of the main text, we obtain the bus frequency $\omega_r/2\pi = 3.007036\pm0.000004$ GHz, the coupling strength 
$g_t/2\pi=17.490\pm0.208$ MHz and transmon decay rate $\gamma_t/2\pi=3.952\pm0.136$ MHz,
yielding a value that reflects the system's damping dynamics and coherence limitations.
Using the above parameters and the theoretical transmission function
\begin{equation}
  S_{21}(\omega_p) \propto \frac{1}{\omega_r - \omega_p - i\kappa/2-\frac{g_t^2}{\omega_q - \omega_p - i\gamma_t/2}},\label{eq:supp-resonator-qubit}
\end{equation}
where $\omega_p$ is the probe frequency, the transmission spectrum can be reconstructed, as shown in Fig.~\ref{fig:transmon_spectra_fit}.

\begin{figure}[hbtp!]
    \centering
    \includegraphics[height=5cm]{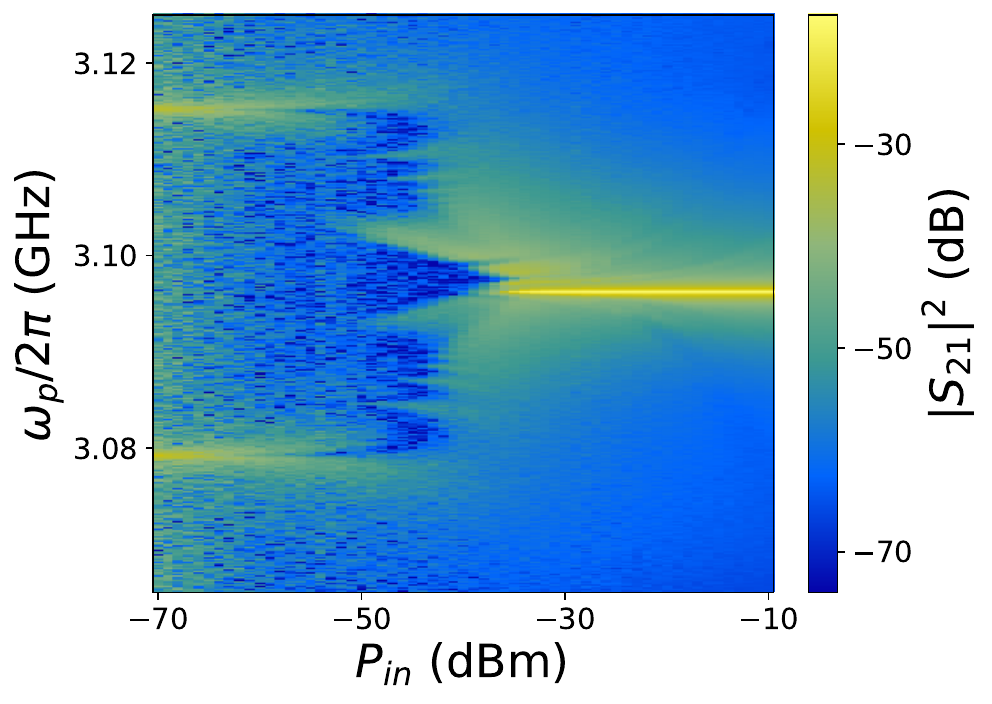}
    \caption{Transmon power sweep, with spectroscopy amplitude (dB) vs probe frequency and VNA output power (dBm).}
    \label{fig:transmon_powersweep}
\end{figure}

We verify the transmon remains in the linear response regime at $-70$ dBm probe power.
Figure~\ref{fig:transmon_powersweep} shows a measured power sweep of the transmon spectroscopy (VNA output power varied from $-70$ to $-10$ dBm).
At the nominal VNA setting used for linear-response measurements ($-70$ dBm), the transmon resonance exhibits only broadening of the line width without the emergence of nonlinear signatures such as supersplitting or additional resolved dressed states~\cite{bishop_2008_nonlinear}.

\subsection{\texorpdfstring{NV$^-$ spin ensemble}{NV- spin ensemble}}

\begin{figure}[hbt]
    \centering
    \includegraphics[height=5cm]{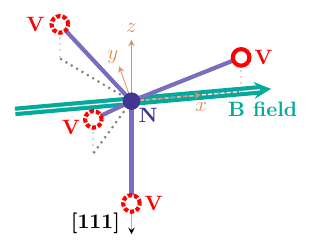}
    \caption{Schematic for orientation of the NV$^{-}$ center with respect to the laboratory axes and applied magnetic field.
}
    \label{fig:NV_structure}
\end{figure}

The diamond is shown as the black square on top of the bus resonator in Fig.~\ref{fig:chip}, and its orientation is illustrated in Fig.~\ref{fig:NV_structure}. With the crystalline $[111]$ axis aligned with the laboratory $z$ axis and the in-plane magnetic field applied along the laboratory $x$ axis, the four possible NV orientations respond differently as the field increases from zero. One subensemble is aligned most closely with the magnetic field and therefore couples strongly at low field. A second pair of subensembles becomes resonant only at higher field values, while the fourth orientation remains perpendicular to the field and does not couple at all. As a result, in our experiment we effectively address only a single NV$^{-}$ subensemble.

\begin{figure}
    \centering
  \includegraphics[height=5cm]{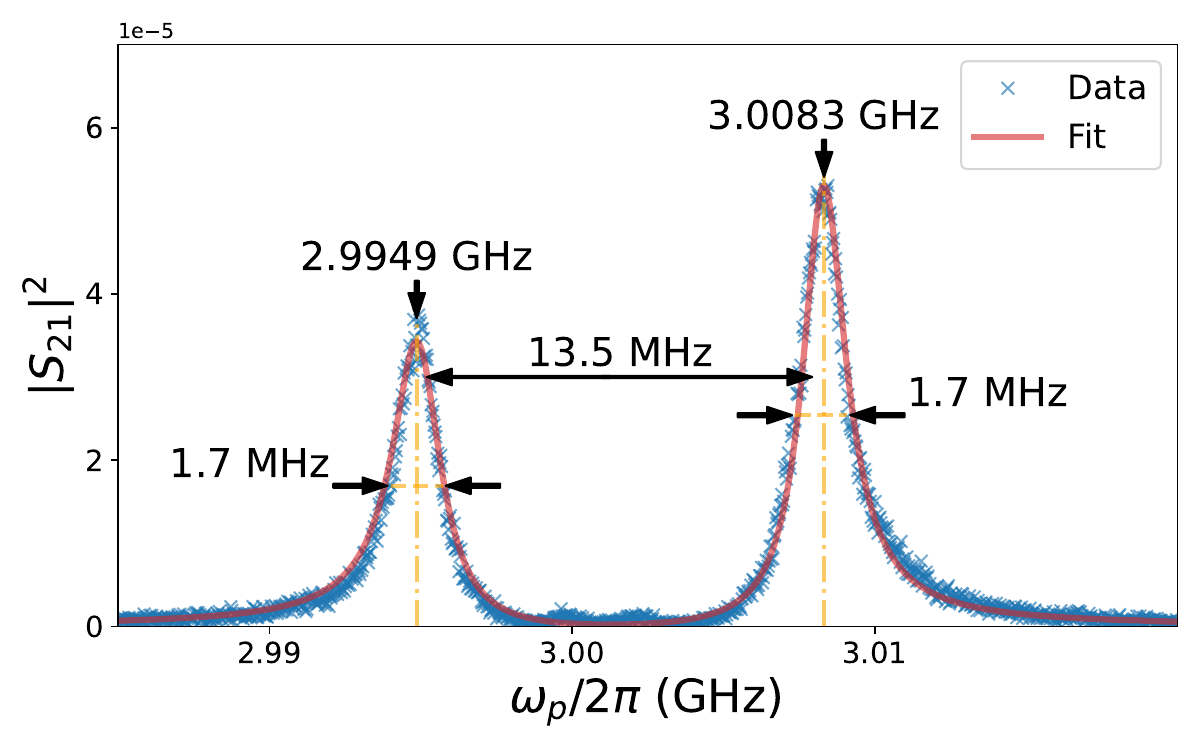}
    \caption{Transmission spectra of the NV$^-$ ensemble coupled to the bus resonator.}
    \label{fig:ensemble_spectra_fit}
\end{figure}

The resonance between the NV$^-$ ensemble and the bus resonator is characterized through spectroscopy, as illustrated in Fig. 1(d) of the main text.
When the ensemble is in resonance with the resonator, two distinct polariton peaks emerge, with a separation of $13.5$ MHz directly yielding the collective coupling strength, see Fig.~\ref{fig:ensemble_spectra_fit}.
The FWHM of these peaks are measured to be $1.7$ MHz for both. 
Employing the relation $\text{FWHM} \approx (\kappa + \Gamma_{\mathrm{ens}})/2$ for the polariton linewidth~\cite{diniz_2011_strongly}, where $\Gamma_{\mathrm{ens}}$ is the spin ensemble decay rate, the contribution of the NV$^-$ ensemble to the dissipation is evaluated as $\Gamma_{\mathrm{ens}}/2\pi \approx 3.2$ MHz.

To fit the data, we consider the transmission spectrum (in the linear-response regime) derived in~\cite{diniz_2011_strongly},
\begin{equation}
\left|S_{21}(\omega_p)\right|^2 \propto \left| \frac{1}{\omega_r -\omega_p - i\kappa/2 -\Omega_e^2\int d\omega \frac{\varrho(\omega)}{\omega-\omega_p-i\gamma_s/2}} \right|^2,
\end{equation}
where $\omega_p$ is the probe frequency, $\kappa$ is the cavity decay rate, $\gamma_s$ is the spin decoherence rate, and $\varrho(\omega)$ is the spectral density profile associated with the NV$^{-}$ ensemble following a $q$-Gaussian distribution~\cite{sandner_2012_strong}:
\begin{equation}
\varrho(\omega) = \rho_0 \left[ 1 - (1 - q) \frac{(\omega - \omega_s)^2}{\Delta^2} \right]^{\frac{1}{1-q}},
\end{equation}
which reduces to Gaussian for $q = 1$ and Lorentzian for $q=2$.
By fitting the experimental data (picking $\gamma_s/2\pi=1$ kHz as a reasonable small value), we obtain the bus frequency $\omega_r/2\pi=3.002001\pm0.000013$ GHz, the ensemble central frequency $\omega_s/2\pi=3.001185\pm0.000020$ GHz, the collective coupling strength $\Omega_e/2\pi = 6.597\pm 0.092$ MHz, the width of $\varrho(\omega)$ as $\Gamma_{\varrho}/2\pi=3.434\pm0.202$ MHz, and $q=1.9591\pm0.0660$.
The fit result is shown in Fig.~\ref{fig:ensemble_spectra_fit}.

\section{Three-mode hybridization theory}

\subsection{The approximated Hamiltonian}

The bus resonator is modeled as a harmonic oscillator with frequency $\omega_r$ and annihilation operator $a$.
The transmon is in general treated as an anharmonic oscillator with fundamental frequency $\omega_t$, anharmonicity parameter $\delta_t$, annihilation operator $b$, and nonlinear term $b^\dagger b^\dagger bb$, which goes beyond the two-level approximation used in the main text and captures the transmon’s intrinsic multilevel nature.
The NV$^-$ ensemble consists of a collection of $N$ spins, which can be viewed as $N$ spin-$1/2$ described by Pauli operators.
The bus-transmon and bus-ensemble couplings are conductive and magnetic dipole couplings, respectively.
The Hamiltonian of this hybrid quantum system can be written as
\begin{equation}
   H = \hbar \omega_a a^\dagger a + \hbar \omega_b b^\dagger b + \hbar \frac{\delta}{2} b^\dagger b^\dagger b b + \hbar \sum_{j=1}^{N}\omega_j \frac{\sigma_j^z}{2} + \hbar g_{t} (a^\dagger b + a b^\dagger) + \hbar\sum_{j=1}^{N} g_j (a^\dagger \sigma_j^- + a \sigma_j^+),
   \label{eq:app-hybrid-ham}
\end{equation}
where we have used the rotating-wave approximation.

In the experiment, the mean number of excitations in the NV$^-$ spin ensemble is much much lower than the number of spins, which allows us to make the Holstein-Primakoff approximation to describe the spin ensemble as a harmonic oscillator.
This approximation is also valid in our case where the spin ensemble has narrow inhomogeneous broadening feature~\cite{diniz_2011_strongly,kurucz_2011_spectroscopic}.
The transmon is simply truncated as a two-level system, as the drive field is weak.
Consequently, Eq.~\eqref{eq:app-hybrid-ham} is approximated as Eq.~(1) in the main text.

\subsection{Linear response regime}

The linear response regime can be approximated as the one-excitation regime to simplify the analysis, where weak driving confines the system dynamics predominantly to single quantum excitations.
Using the single-excitation basis ${|1gG\rangle, |0eG\rangle, |0gE\rangle}$,  the Hamiltonian in Eq.~(1) of the main text can be written as a $3\times3$ matrix:
\begin{equation}\label{eq:Hexcit1}
H_{\text{excit-1}}=\hbar
\begin{pmatrix}
\omega_r & g_t & \Omega_e\\
g_t & \omega_t & 0\\
\Omega_e & 0 & \omega_s
\end{pmatrix}.   
\end{equation}
Analytically solving the eigenenergies and eigenstates of $H_{\text{excit-1}}$ is feasible, as the problem can be transformed to solving a cubic equation.
However, the resulting analytical expressions tend to be intricate and offer limited physical intuition.
For this reason, we instead investigate the following two cases that are directly related to our experiment.

\subsubsection{Triple resonance case}

The resonant case $\omega_r = \omega_t = \omega_s$ gives compact, physically transparent eigenvalues and eigenvectors.
In this case,  
Eq.~\eqref{eq:Hexcit1} can be written as
\begin{equation}
H_{\text{excit-1}} = \hbar
\begin{pmatrix}
\omega_r & g_t & \Omega_e\\
g_t & \omega_r & 0 \\
\Omega_e & 0 & \omega_r
\end{pmatrix},
\end{equation}
which yields eigenenergies
\begin{equation}
E_D=\hbar\omega_r,\qquad E_{\pm}=\hbar\left(\omega_r\pm\Omega_h\right),
\end{equation}
where $\Omega_h=\sqrt{g_t^2 + \Omega_e^2}$. 
The eigenstates can be accordingly solved as
\begin{align}
|1_D\rangle &= |0\rangle|D\rangle, \qquad\qquad\qquad\qquad \text{with } |D\rangle=\frac{\Omega_e|eG\rangle - g_t|gE\rangle}{\Omega_h},\\
|1_\pm\rangle&=\frac{|1gG\rangle\pm|0\rangle|B\rangle}{\sqrt{2}}, \qquad\qquad\text{with } |B\rangle=\frac{g_t|eG\rangle+\Omega_e|gE\rangle}{\Omega_h}.
\end{align}
Therefore, on resonance, one mode is dark (no bus photon) and two bright polariton modes are split by $2\Omega_h$.
This is the emergent three-mode hybridization studied in the experiment.

\subsubsection{Detuned case}

When the transmon frequency is changed while the bus and ensemble are still in resonance, i.e., $\omega_r=\omega_s\neq\omega_t$, we can numerically solve the eigen-problem of $3\times3$ Hamiltonian matrix  
\begin{equation}\hbar
\begin{pmatrix}
\omega_r & g_t & \Omega_e\\
g_t & \omega_t & 0\\
\Omega_e & 0 & \omega_r
\end{pmatrix}  
\end{equation}
to obtain the eigenenergies $E_D(\omega_t), E_\pm(\omega_t)$ and corresponding eigenstates $|1_D(\omega_t)\rangle, |1_\pm(\omega_t)\rangle$, as a function of $\omega_t$.
When probing the system in the linear-response regime, the transmission coefficient $|S_{21}|^2$ attaches maxima near the transition frequencies corresponding to the above eigenenergies.

To understand the transitions between the hybrid modes for this detuned case, we introduce the annihilation operators $B$ and $D$ to $|B\rangle$ and $|D\rangle$, respectively,
\begin{equation}
B=\frac{g_tb+\Omega_eS}{\Omega_h},\qquad D = \frac{\Omega_e b - g_t S}{\Omega_h},
\end{equation}
such that $B^\dagger |0gG\rangle=|B\rangle$ and $D^\dagger |0gG\rangle=|D\rangle$.
Since the inner product $\langle B,D\rangle=0$, we can rewrite the Hamiltonian of Eq.~(1) in the main text as
\begin{equation}
H = \hbar \omega_r a^\dagger a + \hbar \omega_B B^\dagger B + \hbar\omega_D D^\dagger D + \hbar \Omega_h(a^\dagger B+aB^\dagger)+\hbar\chi(B^\dagger D+BD^\dagger)+\dots,
\end{equation}
where $\dots$ denotes the transmon anharmonicity term, which is irrelevant in the one-excitation Hamiltonian, and
\begin{equation}
    \omega_B = \frac{\omega_tg_t^2+\omega_s\Omega_e^2}{\Omega_h^2},\qquad \omega_D = \frac{\omega_t\Omega_e^2+\omega_sg_t^2}{\Omega_h^2},\qquad \chi = \frac{g_t\Omega_e(\omega_t-\omega_s)}{\Omega_h^2}.
\end{equation}
Thus, the bus mode $a$ couples to mode $B$ via the JC interaction, while the mode $D$ only couples to $B$ when $\omega_t\neq\omega_s$.
This matches the observation in the experiment, where an avoided crossing between two bright modes is shown when detuning $\omega_t$ from $\omega_r=\omega_s$, and the dark mode is invisible for the triple resonance case then becoming visible when the detuned $\omega_t$ leads to a nonzero interaction with mode $B$.

\section{Triple resonance and parameter estimation}

\begin{figure}[htbp]
    \centering
    \includegraphics[height=5cm]{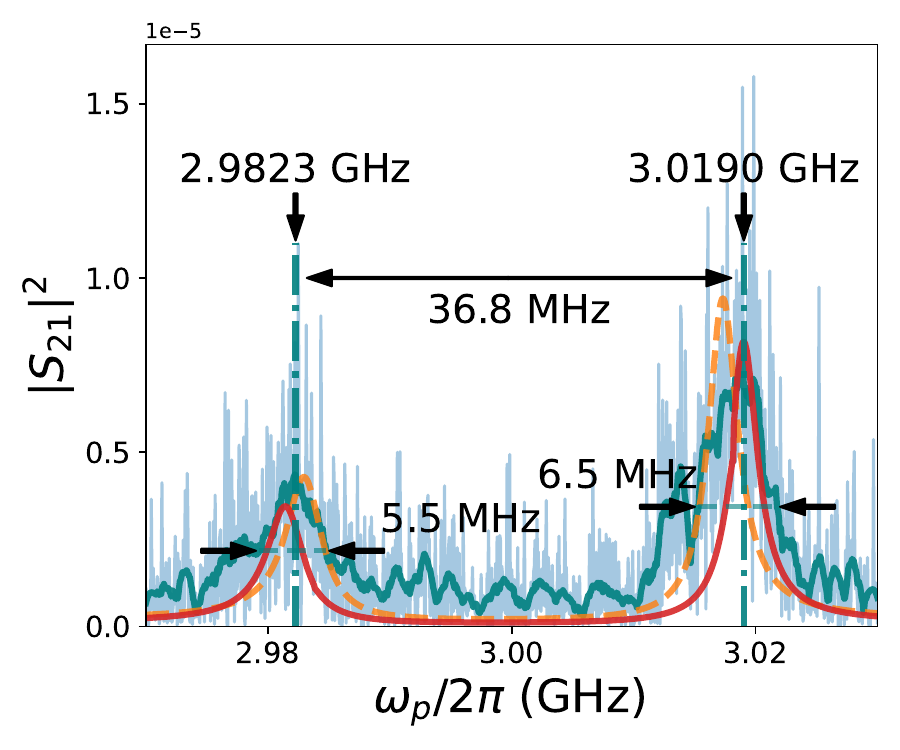}
    \caption{Transmission spectra of the triple resonance condition for the hybrid quantum system. Pale blue: experimental data. Green: processed data for approximating the peaks. Orange: fit of processed data. Red: simulation using parameters from individual system characterization. }
    \label{fig:hybrid-parameters}
\end{figure}

The spectroscopy experiment result of the hybrid quantum system under a weak probe has been illustrated in Fig.~2(a) of the main text.
The triple-resonance spectra with characterized parameters is shown in Fig.~\ref{fig:hybrid-parameters}.
We clearly observe two distinct polariton peaks at $2.9823$ GHz and $3.0190$ GHz, with a splitting of $36.8$ MHz, indicating a collective interaction strength of at least $18.4$ MHz between the transmon and the NV$^-$ ensemble coupled to the bus.
The FWHM of two peaks is measured to be $5.5$ MHz and $6.5$ MHz, respectively, which gives average width of polariton peaks as $(5.5 + 6.5)/2=6.0$ MHz.

\begin{figure}[htbp]
    \centering
    \includegraphics[height=5cm]{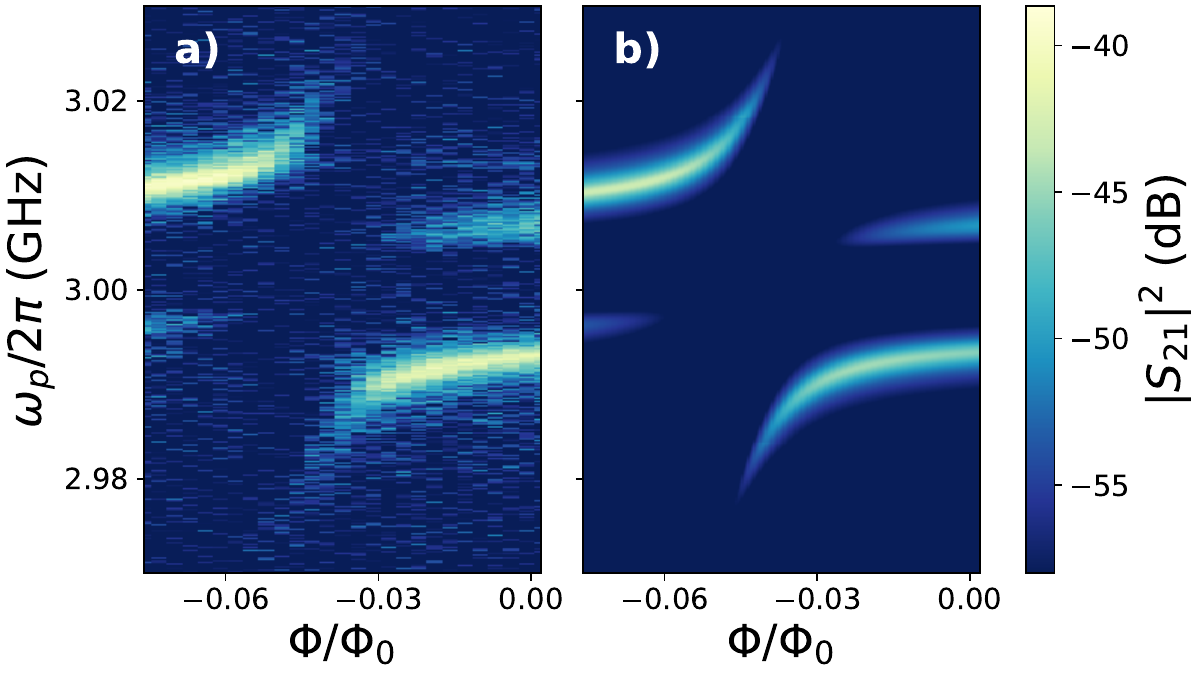}
    \caption{Transmission spectrum of (a) experimental data and (b) simulation.}
    \label{fig:hybrid-simulation}
\end{figure}

The triple-resonance spectra (Fig.~2c of the main text) is expected to be explained by Eq.~(4) of the main text.
However, directly substituting pre-fitted parameters into the model fails to adequately explain the experimental results for several reasons.
First, significant measurement noise obscures the precise signal features.
Second, the resonant frequencies $\omega_r$, $\omega_t$, and $\omega_s$ exhibit drift during data acquisition.
Third, the transmon decay rate $\gamma_t$ is sensitive to magnetic fields, meaning the external field lifts its value.
One may hope to extract all system parameters using a direct global fit of the complete spectrum.
This method turns out impractical as the large parameter space leads to prohibitively long fitting times with poor convergence.
Therefore, to bridge the experimental observations with the theoretical description of Eq.~(4), we employ a stepwise, approximate parameter estimation strategy.
We first estimate the key parameters, including $\omega_r$, $\omega_t$, $\omega_s$, and $\gamma_t$, using the simplified effective Hamiltonian model of Eq.~(1) in the main text.
These values are then combined with the pre-fitted parameters of the NV$^-$ centers to simulate the full spectrum using Eq.~(4).
The results are shown in Fig.~(2) of the main text and Fig.~\ref{fig:hybrid-simulation}.

\section{Full probe-power dependent spectra for nonlinear response}

\begin{figure}[h]
    \centering
    \includegraphics[height=7.5cm]{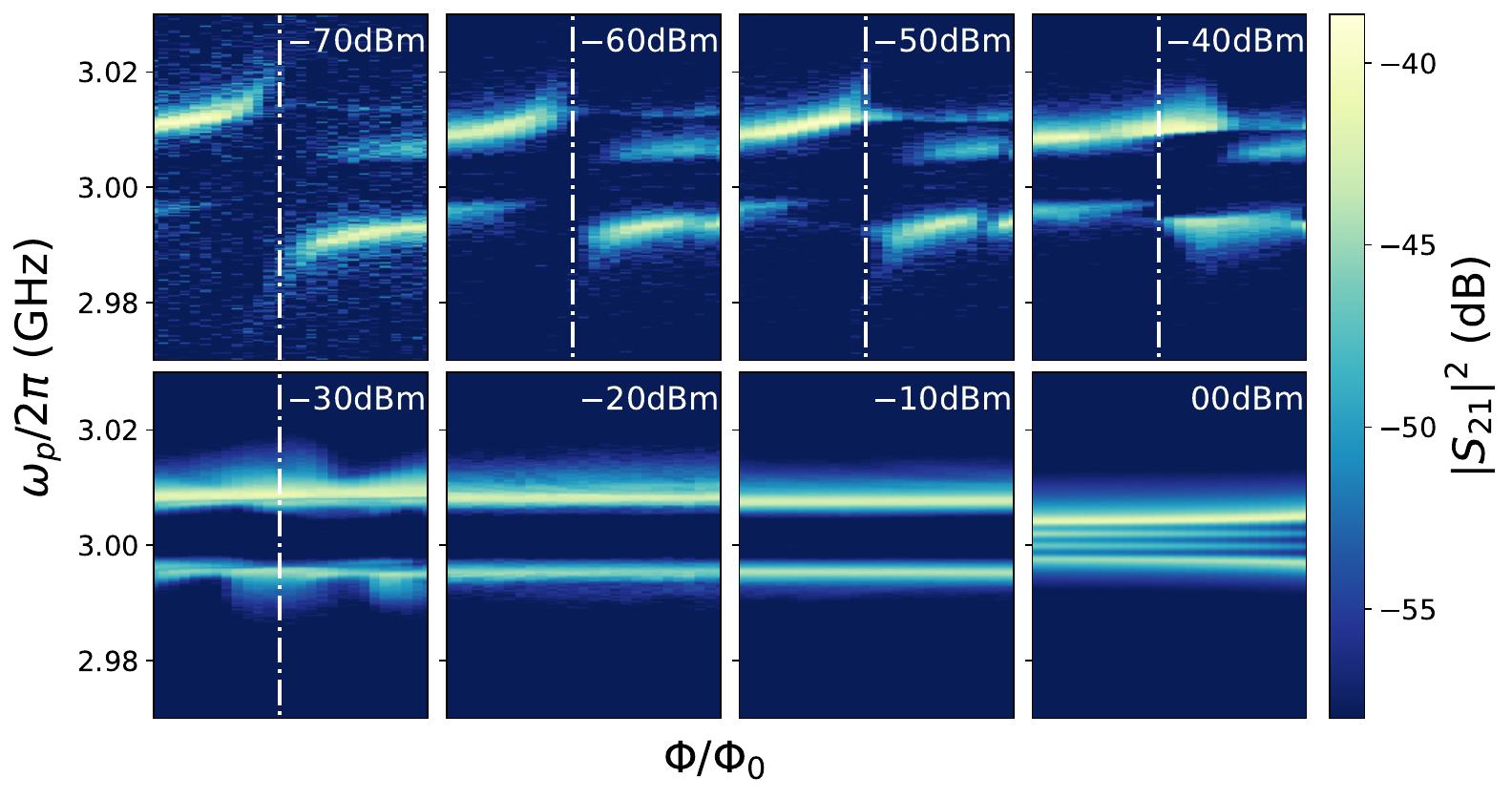}
    \caption{Full probe-power dependent spectra for the hybrid system from $-70$ dBm to $0$ dBm. The dash-dot lines shows the cuts in Fig.~3 of the main text.}
    \label{fig:hybrid_sweep_full}
\end{figure}

Figure ~\ref{fig:hybrid_sweep_full} presents the full probe-power heat maps from which the linecuts shown in Fig. 3 of the main text were extracted.
Each heat map displays the measured transmission coefficient as a function of probe frequency (GHz, vertical axis) and relative flux on transmon (horizontal axis).

\section{Hyperfine-resolved spin model}

The NV$^{-}$ center in diamond exhibits a hyperfine structure due to the interaction between its electron spin ($S=1$) and the nuclear spin of the nitrogen atom ($^{14}$N, $I=1$).
As a result, each electron spin sublevel (e.g., $m_S = 0$ and $m_S = -1$) splits into three hyperfine sublevels, corresponding to the nuclear spin projections $m_I = -1, 0, +1$. 
For transitions between electron spin levels, such as from $m_S = 0$ to $m_S = -1$, the selection rules ($\Delta m_I = 0$) allow transitions only between sublevels with the same $m_I$ value~\cite{yang_2022_observing}.
This leads to three distinct transition frequencies.
Consequently, the NV$^-$ ensemble is actually a collection of three subensembles, with Hamiltonian approximated as
\begin{equation}
   H = \hbar \omega_a a^\dagger a + \hbar \omega_b b^\dagger b + \hbar \frac{\delta}{2} b^\dagger b^\dagger b b + \hbar \sum_{I=0,\pm1}\omega_{s,I}S_I^+S_I^-+ \hbar g_{t} (a^\dagger b + a b^\dagger) + \hbar\sum_{I=0,\pm1} \frac{\Omega_e}{\sqrt{3}} (a^\dagger S_{I}^- + a S_{I}^+).
   \label{eq:app-hybrid-ham-hyper}
\end{equation}
Here, the coupling strength $\Omega_e$ is equally distributed among three subensembles and $\omega_{s,I} = \omega_s + \alpha I$, with $\omega_s=\omega_r$, and $\alpha=2.2$ MHz.

\begin{figure}[h]
    \centering
    \includegraphics[height=5cm]{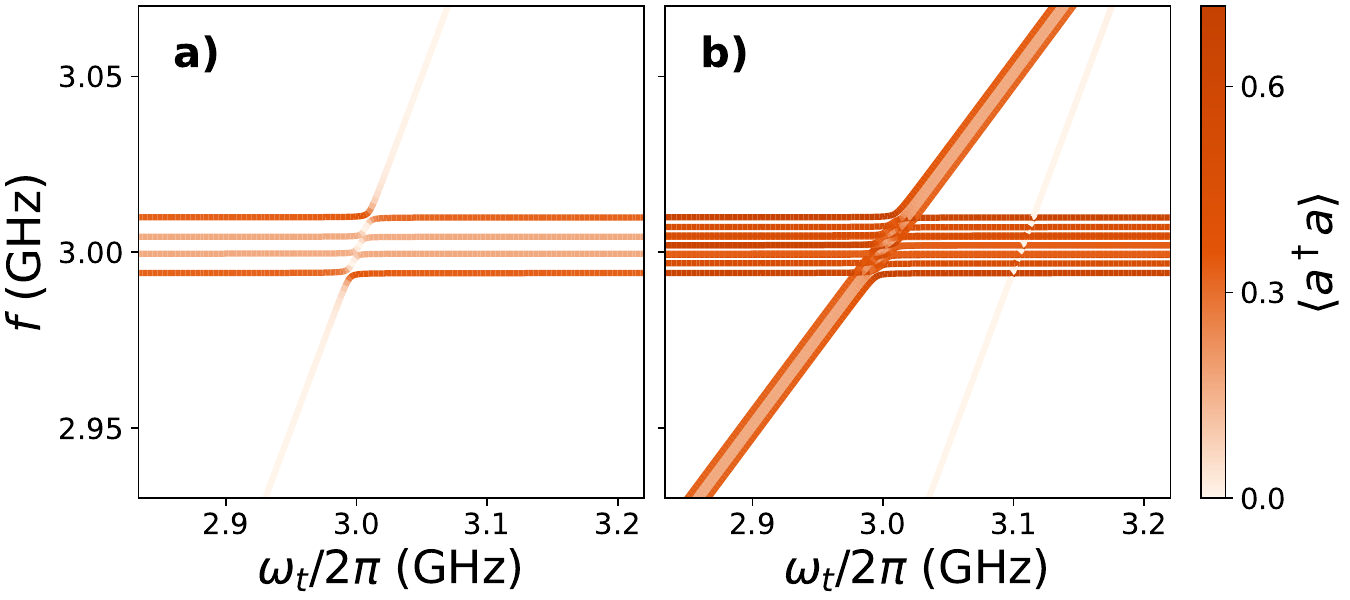}
    \caption{(a) One-photon and (b) two-photon transitions of the strongly coupled hybrid quantum system.}
    \label{fig:supp-N14}
\end{figure}

To interpret Fig.~4 in the main text, we numerically solve the eigenvalue problem of this Hamiltonian within the manifold of no more than two excitations, using determined parameters from experimental data.
The eigenstates are dressed states of resonator mode, transmon mode, and three collective modes of subensembles, and can be classified as two sets with one total excitation and two total excitations respectively.
Figure~\ref{fig:supp-N14}(a) shows energies of the one-excitation dressed states changing with detuned transmon frequency, which exhibits two avoided crossing lines and three bright-dark-bright lines indicating that transmon exchanging energy with the composed resonator and three subensemble system.
These lines constitutes the main pattern in Fig.~4 of the main text.
Meanwhile, Fig.~\ref{fig:supp-N14}(b) shows half of the energies in the two-excitation dressed states.
These transitions serve as weak two-photon transitions observed in Fig.~4 of the main text.

\bibliography{SpectroHQS-Exp}